\documentclass[a4paper,11pt]{article}
\usepackage[reqno,fleqn]{amsmath}
\usepackage{color,graphicx}
\usepackage[T1]{fontenc}
\author{Elnaz Amirkhanlou\footnote{eliamirkhanlou@yahoo.com}, Behnam Mohammadi\footnote{be.mohammadi@urmia.ac.ir}\\
Department of Physics, Urmia University, Urmia, Iran}
\title{Amplitude analysis and branching fraction calculations of $B_c$ meson decays into $B_s\pi$, $DK\pi$ and $D_sKK$ final state mesons}
\begin{document}
\maketitle
\begin{abstract}

In this study, several newly observed decay modes of the $B_c^+$
meson of the form $B_c^+\rightarrow Dh^+h^-$, where $D$ denotes a
charmed meson and $h^{\pm}$ represents a charged pion or kaon,
have been investigated. The decay channels $B_c^+\rightarrow
D^{+}K^+\pi^{-}$, $B_c^+\rightarrow D^{*+}K^+\pi^{-}$ and
$B_c^+\rightarrow D_s^+K^+K^-$ were observed for the first time.
The LHCb Collaboration has measured their branching fractions
relative to the reference decay $B_c^+\rightarrow B_s^0\pi^+$, and
reported experimental values including
$\mathcal{R}^{\exp}(B_c^+\rightarrow
D^{+}K^+\pi^{-})=(1.96\pm0.23\pm0.08\pm0.10)\times 10^{-3}$ ,
$\mathcal{R}^{\exp}(B_c^+\rightarrow
D^{*+}K^+\pi^{-})=(3.67\pm0.55\pm0.24\pm0.20)\times10^{-3}$ and
$\mathcal{R}^{\exp}(B_c^+\rightarrow
D_s^+K^+K^{-})=(1.61\pm0.35\pm0.13\pm0.07)\times 10^{-3}$. In this
regard, by performing precise calculations using the factorization
approach, we have computed similar branching ratio values,
including $\mathcal{R}(B_c^+\rightarrow
D^{+}K^+\pi^{-})=(1.70\pm0.08)\times 10^{-3}$,
$\mathcal{R}(B_c^+\rightarrow
D^{*+}K^+\pi^{-})=(3.63\pm0.45)\times 10^{-3}$ and
$\mathcal{R}(B_c^+\rightarrow D_s^+K^+K^{-})=(1.09\pm0.08)\times
10^{-3}$. The obtained theoretical results show good consistency
with the recent experimental measurements reported by the LHCb
collaboration, confirming the effectiveness of the factorization
formalism in describing these $B_c^+$ decay modes.
\end{abstract}

\section{Introduction}
Recently, extensive studies have been devoted to searching for new
physics in low-energy flavor dynamics, particularly at
$B$-factories. Unlike the purely beauty or charm hadrons, the
decay mechanisms of the $B_c$ meson involve both $b\rightarrow c$
and $\bar{c}\rightarrow\bar{s},\bar{d}$ transitions. Consequently,
they provide access to various elements of the
Cabibbo-Kobayashi-Maskawa (CKM) matrix and serve as a sensitive
probe for testing the hypotheses of factorization and heavy-quark
spin symmetry. Non-leptonic $B_c$ transitions into charmonium final states have been investigated within model-independent
frameworks based on heavy-quark expansions \cite{N.L1}.\\
Experimental measurements in flavor physics, specifically concerning the $B_c$ meson, have experienced a
decisive leap forward in recent years. The simultaneous observations by CMS \cite{CMS1} and LHCb \cite{LHCB1},
confirming and precisely determining the mass of the excited $B_c$ state, have provided compelling evidence for
the validity of quark-model predictions.\\
These achievements have now paved the way for more refined explorations, and with LHCb taking the lead in 2025
\cite{LHCB2}, \cite{LHCB3} to observe the orbitally excited $B_c(1P)$ states, research in this domain is transitioning
from the discovery phase to the precision measurement era of spectral parameters.\\
At this stage, the focus has shifted to the dynamical mechanisms governing the production and decay of the $B_c$ meson.
In this context, precise measurements of newly observed decay rates, such as the three-body decays recently reported by LHCb,
serve as critical testing grounds for theoretical frameworks describing strong interactions in heavy-quarkonium systems. Accordingly,
constructing a consistent theoretical framework capable of describing the dynamics of three-body decays, based on the factorization
expansion of the strong interaction, has become an essential undertaking. These unprecedentedly accurate data now enable stringent
tests of theoretical predictions. Furthermore, recent precision measurements of mass \cite{LHCB4}, lifetime \cite{LHCB5},
production fractions \cite{LHCB6}, and various decay modes have significantly deepened our understanding of this unique
two-heavy-quark system \cite{LHCB7}.\\
From a theoretical standpoint, the properties of the $B_c$ meson
have been predicted using a variety of frameworks, including
non-relativistic and relativistic quark-potential models
\cite{Q.L1}, lattice QCD (Quantum Chromodynamics)
calculations \cite{C.T.H1}, and effective-field-theory approaches \cite{E.J1}.\\
These studies have offered precise predictions for the
ground-state mass as well as a rich spectrum of excited $S$-,
$P$-, $D$-, and higher orbital states \cite{X.J1}, \cite{T.Y1}. On
the experimental side, the remarkable advances achieved over the
past two decades by detectors such as CDF, LHCb, ATLAS, and CMS
have opened a new era in $B_c$ spectroscopy. Despite these
significant
developments, crucial gaps remain.\\
First, the uncertainties in the parameters (mass and width) of
many predicted excited states, particularly those near the current
experimental discovery threshold, necessitate more refined
theoretical calculations and targeted experimental analyses.
Second, the detailed mechanisms governing the decays of excited
$B_c$ states, especially hadronic and radiative transitions,
require further investigation. Consequently, there is a pressing
need for a comprehensive and systematic analysis that confronts
the predictions of various theoretical models with the most recent
experimental data, including the latest LHCb results \cite{LHCB8},
to achieve a coherent understanding of the $B_c$ spectrum.\\
This progress directly informs the study of the $B_c$ meson, a
unique laboratory for testing QCD potentials and weak decay
mechanisms due to its distinct heavy-quark flavor composition
\cite{M.S1}. While its ground-state properties are well-measured
\cite{CDF1}, \cite{V.M1}, \cite{CDF2}, detailed investigations of
its excited states and their decay pathways, including various
hadronic and
semi-leptonic transitions, remain central to ongoing research.\\
Following our initial consideration of the broader landscape of
heavy hadron decays and emphasis on the fundamental significance
of studying the $B_c^+$ meson, this section introduces the
specific focus of the present investigation. Our primary attention
is directed toward the newly observed $B_c^+$ decay channels
belonging to the general form $B_c^+\rightarrow D h^+h^-$, where
$D$
denotes a charm meson and $h^\pm$ represents a charged pion or kaon \cite{LHCB8}.\\
This study specifically covers the channels recently identified by
the LHCb collaboration: $B_c^+ \rightarrow D^+K^+\pi^-$,
$B_c^+\rightarrow D^{*+}K^+\pi^-$, and $B_c^+\rightarrow
D_s^+K^+K^-$. LHCb reported the experimental ratios for these
decays relative to the reference decay $B_c^+ \to B_s^0 \pi^+$ as:
$\mathcal{R}^{\exp}(B_c^+\rightarrow D^+ K^+
\pi^-)=(1.96\pm0.23\pm0.08\pm0.10)\times 10^{-3}$,
$\mathcal{R}^{\exp}(B_c^+\rightarrow
D^{*+}K^+\pi^-)=(3.67\pm0.55\pm0.24\pm0.20)\times 10^{-3}$ and
$\mathcal{R}^{\exp}(B_c^+\rightarrow D_s^+K^+K^-)=(1.61\pm0.35\pm0.13\pm0.07) \times 10^{-3}$ \cite{LHCB8}.\\
To validate these experimental observations and deepen our
understanding of the underlying dynamics, we have performed
parallel theoretical calculations utilizing the rigorous
factorization approach. Our computed results, demonstrating high
computational precision, are: $\mathcal{R}(B_c^+ \to D^+ K^+
\pi^-) = (1.70\pm0.08)\times 10^{-3}$,
$\mathcal{R}(B_c^+\rightarrow D^{*+} K^+\pi^-)
=(3.63\pm0.45)\times 10^{-3}$ and
$\mathcal{R}(B_c^+\rightarrow D_s^+ K^+ K^-)=(1.09\pm0.08)\times 10^{-3}$.\\
These theoretical findings show excellent agreement with the
recent LHCb measurements, thereby confirming the efficacy of the
factorization formalism in describing these complex $B_c^+$ decay
modes and providing a foundation for more precise tests of QCD
models in the heavy-hadron regime.

\section{The $B^+_c\rightarrow B^0_s\pi^+$ decay}

\begin{figure}[t]
\begin{center} \includegraphics[scale=1]{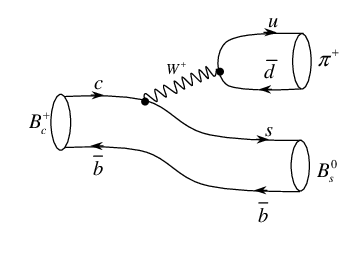}
\caption{\label{fig1}The current-current diagram contributing to
$B^+_c\rightarrow B^0_s\pi^+$ decay.}
\end{center}
\end{figure}

To calculate the decay of the $B_c^+$ meson, we employ a
theoretical framework based on the factorization approach and the
Wilson coefficients associated with the effective four fermion
operators. Within this framework, the hadronic matrix elements
can be parametrized in terms of transition form factors and decay constants.\\
In this work, we consider the two-body decay $B_c^+\rightarrow
B_s^0\pi^+$. The $B_c^+$ meson is a heavy meson composed of two
heavy quarks, namely the $\bar{b}$ and $c$ quarks, while the
$B_s^0$ meson consists of the $\bar{b}$ and $s$ quarks. The study
of such heavy-meson decays provides important information about
weak interactions and hadronic dynamics
within the standard model (SM).\\
For the decay examined in this work, $B_c^+\rightarrow
B_s^0\pi^+$, the dominant contribution arises from the
current-current weak interaction in which the charm quark
undergoes the transition $c\rightarrow su\bar{d}$, while the
$\bar{b}$ quark acts as a spectator. The corresponding Feynman
diagram for this process is shown in Fig. \ref{fig1}.
Within the factorization framework, the decay amplitude is
decomposed into the product of two independent hadronic matrix
elements: the $B_c^+ \rightarrow B_s^0$ transition matrix element,
and the vacuum-to-pion matrix element, which is parametrized by
the pion decay constant $f_\pi$. \textcolor{red}{The effective
weak Hamiltonian for the $B_c^+ \to B_s^0 \pi^+$ decay, which is
mediated by the $c \to s \bar{d} u$ weak transition, is given by:
\begin{equation}\label{eq1}
\mathcal{H}_{\text{eff}}(B_c^+ \to B_s^0 \pi^+)
=\frac{G_F}{\sqrt{2}} V_{cs} V_{ud}^* \left[ c_1 (\bar{s}_n
c_n)_{V-A} (\bar{u}_m d_m)_{V-A} + c_2 (\bar{s}_n c_m)_{V-A}
(\bar{u}_m d_n)_{V-A} \right],
\end{equation}{\setlength\arraycolsep{.75pt}
where $G_F$ is the Fermi coupling constant, $n, m$ denote the
color indices, and the current operator is defined as $(\bar{q}_1
q_2)_{V-A} = \bar{q}_1 \gamma^\mu(1-\gamma_5)q_2$. Under the naive
factorization hypothesis, the hadronic matrix element is
approximated as the product of two independent bilinear currents
\cite{B.Mo}:
\begin{eqnarray}\label{eq2}
\mathcal{A}_{B^+_c\rightarrow B^0_s\pi^+}&=& \langle B_s^0 \pi^+| \mathcal{H}_{\text{eff}} | B_c^+ \rangle \nonumber\\
&=&\frac{G_F}{\sqrt{2}}V_{cs}V_{ud}^*a_1\left\langle\pi^+|\bar{u}\gamma^\mu(1-\gamma_5)d|0\right\rangle\left\langle
B^0_s|\bar{s}\gamma_\mu(1-\gamma_5)c|B_c^{+}\right\rangle\\
\nonumber &=& \frac{G_F}{\sqrt{2}}V_{cs}V_{ud}^*a_1
f_{\pi}(m_{B_c}^2-m_{B_s}^2)F_0^{B_c\rightarrow B_s}(m^2_{\pi}),
\end{eqnarray}}}
here $f_{\pi}$ is decay constant and the form factor is obtained from the following equation \cite{RDC}
{\setlength\arraycolsep{.75pt}
\begin{eqnarray}\label{eq3}
F_0^{B_c\rightarrow B_s}(q^2)&=&\frac{f(0)}{1-q^2/m^2_{fit}+\sigma q^4/m^4_{fit}},
\end{eqnarray}
where the $f(0)=0.73\pm0.03$, $\sigma=0.60\pm0.18$ and
$m_{fit}=1.77\pm0.22$
\cite{RDC}.
The quantities $a_i$ (i=1,2) are evaluated with the Wilson
coefficients in the next-to-leading logarithmic order
\cite{M.Be1}.
\begin{eqnarray}\label{eq4}
a_{2i-1}=c_{2i-1}+\frac{1}{3}c_{2i},\quad
a_{2i}=c_{2i}+\frac{1}{3}c_{2i-1},\quad i=1,2.
\end{eqnarray}
The Wilson coefficients $c_i$ are used at the three deferent
scales obtained from the naive dimensional regularization scheme
as listed in Tab. \ref{tab1}.

\begin{table}[t]
\centering\caption{\label{tab1}  Wilson coefficients $c_i$ in the
naive dimensional regularization scheme \cite{M.Be1}.}
\begin{tabular}{ccc}
$\mu$ & $c_1$ & $c_2$ \\[2pt]\hline\hline
$m_b/2$ & 1.137 & $-0.295$ \\[2pt]
$m_b$ & 1.081 & $-0.190$ \\[2pt]
$2m_b$ & 1.045 & $-0.113$ \\ \hline
\end{tabular}
\end{table}

The CKM matrix elements used in this work are \cite{G.K1}
{\setlength\arraycolsep{.75pt}
\begin{eqnarray}\label{eq5}
 |V_{ud}|&=&0.97367\pm0.00032,\quad |V_{cb}|=(41.1\pm1.2)\times10^{-3},\quad |V_{ub}|=(3.82\pm0.20)\times10^{-3},\nonumber\\
|V_{cs}|&=&0.975\pm0.006,\qquad\quad |V_{us}|=0.22431\pm0.00085.
\end{eqnarray}}
The branching fractions of $B^+_c\rightarrow B^0_s\pi^+$ decay is given by
\begin{eqnarray} \label{eq6}
\mathcal{B}r(B^+_c\rightarrow B^0_s\pi^+)=\frac{p_c}{8\pi
m^2_{B_c}\Gamma_{tot}}|\mathcal{A}_{B^+_c\rightarrow B^0_s\pi^+}|^2,
\end{eqnarray}
where $\Gamma_{tot}$ for $B^+_c$
is $(1.29\pm0.01)\times10^{-12}$ GeV \cite{PDG} and
{\setlength\arraycolsep{.75pt}
\begin{eqnarray}\label{eq7}
p_c=\frac{\sqrt{(m^2_{B^+_c}-(m_{B_s}+m_{\pi^+})^2)(m^2_{B^+_c}-(m_{B_s}-m_{\pi^+})^2)}}{2m_{B_c}}.
\end{eqnarray}}

\section{The amplitudes and branching fractions of $B^+_c\rightarrow D^+K^+\pi^-$, $B^+_c\rightarrow D^{*+}K^+\pi^-$ and
$B^+_c\rightarrow D^+_sK^+K^-$ decays}

To evaluate the amplitudes and branching fractions of the
three-body decays $B_c^{+}\rightarrow D^{+}K^{+}\pi^{-}$,
$B_c^{+}\rightarrow D^{*+}K^{+}\pi^{-}$ and $B_c^{+}\rightarrow
D_s^{+}K^{+}K^{-}$, we employ the effective weak Hamiltonian
describing the $b\rightarrow c\bar{u}s$ and $c\rightarrow s u\bar{s}$ transitions.\\
The dominant contributions arise from the color-allowed external
$W$-emission topology. In this class of diagrams, the charm quark
inside the $B_c^{+}$ meson undergoes the transition $c\rightarrow
s$ together with the emission of a virtual $W^-$ boson, which
subsequently materializes into the final light meson ($K^{+}$ or
$\pi^{-}$). The remaining quark system hadronizes to form the
$D^{+}$, $D^{*+}$, or $D_s^{+}$ meson.
These diagrams provide the leading factorizable contributions.\\
Through detailed analysis of three-body decay processes, one can
extract important physical observables such as decay amplitudes,
branching fractions and kinematical distributions, which provide
valuable information on the underlying strong-interaction dynamics
described by QCD. In particular, three body decays offer a rich
laboratory for studying hadronic dynamics through techniques such
as the Dalitz-plot analysis, which allows one to explore the
interplay
between resonance and non-resonant contributions \cite{E.A8}.\\
In this section, we derive the decay amplitudes for the processes
$B_c^{+}\rightarrow D^{+}K^{+}\pi^{-}$, $B_c^{+}\rightarrow
D^{*+}K^{+}\pi^{-}$, and $B_c^{+}\rightarrow D_s^{+}K^{+}K^{-}$
based on the factorization framework and the dominant
color-allowed external $W$-emission diagrams shown in Fig.
\ref{fig2}.
\begin{figure}[t]
\begin{center} \includegraphics[scale=0.7]{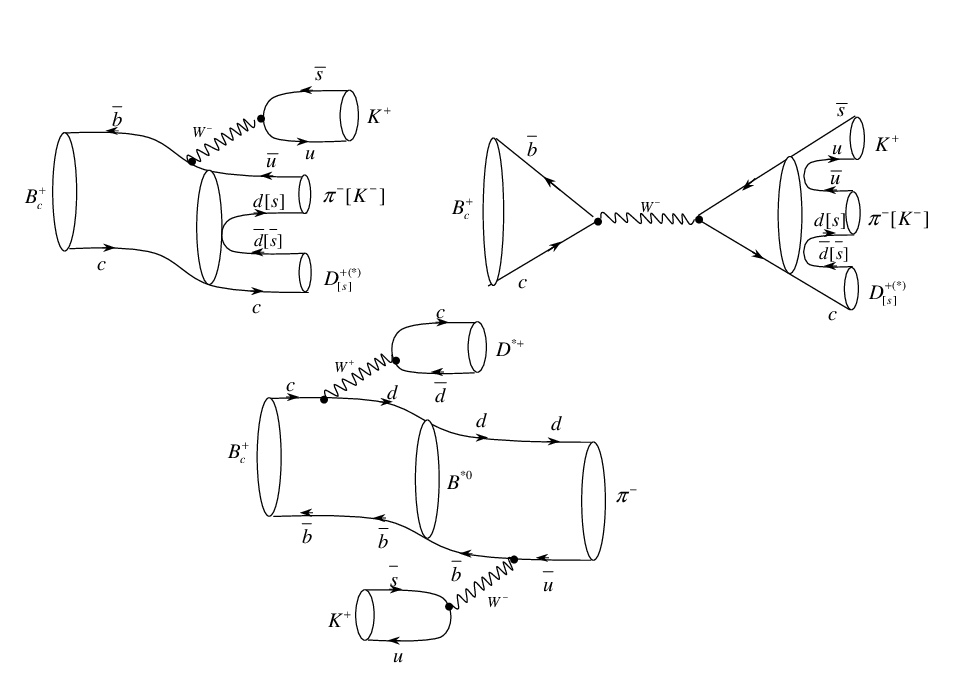}
\caption{\label{fig2}Feynman diagrams current-current and annihilation topologies contributing to
the decays $B_c^{+}\rightarrow D^{+}K^{+}\pi^{-}$, $B_c^{+}\rightarrow D^{*+}K^{+}\pi^{-}$, and
$B_c^{+}\rightarrow D_s^{+}K^{+}K^{-}$ as well as the pole diagram for the $B_c^{+}\rightarrow D^{*+}K^{+}\pi^{-}$ decay.}
\end{center}
\end{figure}
\textcolor{red}{The non-resonant decay $B_c^{+} \rightarrow
K^{+}\pi^{-} D^{+(*)}_{[s]}$ receives contributions from two
distinct quark-level weak transitions: $b \rightarrow u \bar{u} s$
and $b \rightarrow c \bar{c} s$. The corresponding effective weak
Hamiltonians are written as:
\begin{equation}\label{eq8}
\mathcal{H}_{\text{eff}}^{(1)} = \frac{G_F}{\sqrt{2}} V_{ub}
V_{us}^*\sum_{i} C_i O_i, \quad \mathcal{H}_{\text{eff}}^{(2)} =
\frac{G_F}{\sqrt{2}} V_{cb} V_{cs}^* \sum_{i} C_i O_i,
\end{equation}
where $O_i$ are the local four-quark operators. By applying the
factorization ansatz to the transition matrix elements, the
non-resonant amplitude is expressed as:}
{\setlength\arraycolsep{.75pt}
\begin{eqnarray}\label{eq9}
&&\mathcal{A}^{NR}(B_c^{+}(p_B)\rightarrow
K^{+}(p_1)\pi^{-}[K^{-}](p_2)D^{+(*)}_{[s]}(p_3))=i\frac{G_F}{\sqrt{2}}\Big[\Big(V_{ub}V_{us}^{*}
a_1 \nonumber\\&&
\left\langle\pi^{-}[K^{-}](p_2)D^{+(*)}_{[s]}(p_3)|\textcolor{red}{\bar{u}\gamma^\mu(1-\gamma_5)b}|B_c^{+}(p_B)
\right\rangle\left\langle
K^{+}(p_1)|\textcolor{red}{\bar{s}\gamma_\mu(1-\gamma_5)u}|0\right\rangle\Big)\nonumber\\
&& +\Big(V_{cb}V_{cs}^{*} a_2 \left\langle
K^{+}(p_1)\pi^{-}[K^{-}](p_2)D^{+(*)}_{[s]}(p_3)|\textcolor{red}{\bar{c}\gamma^\mu(1-\gamma_5)b}|0
\right\rangle\left\langle
B_c^{+}(p_B)|\textcolor{red}{\bar{s}\gamma_\mu(1-\gamma_5)c}|0
\right\rangle\Big)\Big].
\end{eqnarray}}
In $B_c^{+}\rightarrow D^{*+}K^{+}\pi^{-}$ decay,
the hadronic matrix elements appearing in the factorized amplitude are written as
 products of separate matrix elements, each of which is evaluated individually.
 The decay constant is defined through the matrix
 element $\langle
K^+(p_1)|\textcolor{red}{\bar{s}\gamma_\mu(1-\gamma_5)u}|0\rangle=if_Kp_1^\mu$
\cite{A.G}. We then evaluate the remaining matrix elements in
terms of the relevant transition form factors. The form factor
enters through the matrix element
$\langle\pi^-D^{*+}|\textcolor{red}{\bar{c}\gamma^\mu(1-\gamma_5)b}|B^+_c\rangle$.
Since the $D^{*+}$ meson is not a light meson, this matrix element
is evaluated by considering the contribution of a virtual
intermediate state, implemented through the $B^{*0}$ pole shown in
Fig. \ref{fig2}. The pole contribution is described in terms of
two strong and weak vertices as \cite{C.A1}
\begin{eqnarray}\label{eq10}
\mathcal{A}^{\mu}_{B^*K^+\pi^-}\frac{i(-g_{\mu\nu}+p_{B^*\mu}p_{B^*\nu}/m^2_{B^*})}
{p^2_{B^*}-m^2_{B^*}}\mathcal{A}^{\nu}_{B^+_cB^*D^{*+}}.
\end{eqnarray}
The strong vertex is achieved via \cite{C.A1}
\begin{eqnarray}\label{eq11}
\epsilon_{B^*\nu}\mathcal{A}^{\nu}_{B^+_cB^*D^{*+}}=\langle
B^*(p_{B^*},\epsilon_{B^*})D^{*+}(p_3,\epsilon_3)|B^+(p_B)
\rangle=i\sqrt{2}g_{B^+_cB^*D^{*+}}\epsilon_{\mu\nu\alpha\beta}\epsilon^{\mu}\epsilon_{\nu}p_{B_c}^{\alpha}p_3^{\beta},
\end{eqnarray}
where
\begin{eqnarray}\label{eq12}
g_{B^+_cB^*D^{*+}}={\frac{2g\sqrt{m_Bm_{B^*}}}{f_{D^*}}}.
\end{eqnarray}
\textcolor{red}{Here, $p_{B^*}=p_B-p_1$. The effective strong coupling $g$ has been
estimated in several phenomenological studies, with the values
$g=0.59\pm0.01\pm0.07$ \cite{A.A11},
$g=0.50\pm0.10$ \cite{DX}, and $g=0.30\pm0.10$ \cite{Y.Mo}. In the numerical analysis, we adopt
$g=0.30$ as the reference value following Ref. \cite{Y.Mo}, and vary
$g$ within the above phenomenological range to assess the model
dependence. The resulting spread is included in the quoted theoretical
uncertainties}. $m_B$ and
$m_{B^*}$ denote the masses of the $B$ and $B^*$ mesons,
respectively; and $f_{D^*}$ is the $D^{*+}$ decay constant. The
weak vertex is \cite{C.A1}
\begin{eqnarray}\label{eq13}
\epsilon_{B^*\mu}\mathcal{A}^{\mu}_{B^*K^+\pi^-}=\frac{iG_F}{\sqrt{2}}a_1V_{ub}V^*_{us}\langle
\pi^-(p_2)|\textcolor{red}{\bar{u}\gamma^\mu(1-\gamma_5)b}|B^{*}(p_{B^*},\epsilon_{B^*})\rangle
\langle
K^+(p_1)|\textcolor{red}{\bar{s}\gamma_\mu(1-\gamma_5)u}|0\rangle.
\end{eqnarray}
For the matrix element $\langle
\pi^-(p_2)|\textcolor{red}{\bar{u}\gamma^\mu(1-\gamma_5)b}|B^{*}(p_{B^*},\epsilon_{B^*})\rangle$, we have \cite{D.A1}
\begin{eqnarray}\label{eq14}
\langle
\pi^-(p_2)|\textcolor{red}{\bar{u}\gamma^\mu(1-\gamma_5)b}|B^{*}(p_{B^*},\epsilon_{B^*})\rangle
&=&T_1i\epsilon_{\mu\nu\alpha\beta}\epsilon_{B^*}^{\nu}p_{B^*}^{\alpha}p_2^{\beta}-T_2m^2_{B^*}\epsilon_{{B^*}\mu}-T_3(\epsilon_{B^*}\cdot
p_2)(p_{B^*}+p_2)\nonumber\\
&-&T_4(\epsilon_{B^*}\cdot p_2)(p_{B^*}-p_1)_{\mu}
\end{eqnarray}
The form factors $T_i$s are defined as follows \cite{E.A7}
\begin{eqnarray}\label{eq15}
&&T_1=-\frac{f_+-f_-}{m_{B_c}},\quad
T_2=\frac{1}{m^2_{B_c}}\Big((f_++f_-)m_{B_c}+(f_+-f_-)\frac{p_{B^*}\cdot
p_1}{m_{B_c}}\Big),
\quad T_3=T_4=-\frac{f_+-f_-}{2m_{B_c}}\nonumber\\
&&f_{\pm}=-\frac{1}{4\sqrt{m_{B_c}m_2}}(m_{B_c}\mp
m_2)\xi(\omega),
\end{eqnarray}
where $\xi(\omega)=1-\rho^2(\omega-1)$, $\rho^2=1.2\pm0.2$ and
$\omega=(m^2_{B_c}+m^2_2-u)/(2m_{B_c}m_2)$ \cite{R.D2}. Particle
interactions are commonly described using Lorentz invariant
variables. In two-body decay, the momentum of the decay products
in the rest frame of the parent particle is determined by
the masses of the parent ($m_{B_c}$) and decay products.\\
In contrast, three-body decay introduces two additional degrees of
freedom, as the three final-state particles ($m_1, m_2, m_3$) can
be paired in three different ways, resulting in invariant masses
$m_{12}$, $m_{13}$, and $m_{23}$. We define these as $m_{12}^2=s$,
$m_{23}^2=t$, and $m_{31}^2=u$ (where \textcolor{red}{$m_{ij}^2 = (p_i + p_j)^2$} and
$c^2=1$). In this case, we have $s=(p_1+p_2)^2=(p_{B_c}-p_3)^2$,
$t=(p_2+p_3)^2=(p_{B_c}-p_1)^2$ and
$u=(p_3+p_1)^2=(p_{B_c}-p_2)^2$ \cite{D.B2}. Conservation of
four-momentum gives us the relation
$s+t+u=m_{B_c}^2+m_1^2+m_2^2+m_3^2$ \cite{D.B2}.\\
By substituting Eqs. (\ref{eq11}), (\ref{eq12}) and (\ref{eq13})
into Eq. (\ref{eq10}) and using the energy-momentum relations and
the four polarization vectors, the factorized expression for the
pole contribution to the amplitude can be written as:
{\setlength\arraycolsep{.75pt}
\begin{eqnarray}\label{eq16}
\langle\pi^-D^{*+}|\textcolor{red}{\bar{c}\gamma^\mu(1-\gamma_5)b}|B^+_c\rangle
&\langle&
K^+|\textcolor{red}{\bar{s}\gamma_\mu(1-\gamma_5)u}|0\rangle=i\sqrt{2}f_K\frac{g_{B^+_cB^*D^{*+}}}{s-m^2_{B^*}}
T_1\Big[\Big(\frac{m_{B_c}^2}{2m_3}(t-m^2_2-m^2_3)\Big)\nonumber\\
&+&\Big(\frac{m_2^2}{2m_3}(s+m^2_3-m^2_{B_c})\Big)-\Big(\frac{1}{2m_3}(s-m^2_3-m^2_{B_c})(m^2_2+m^2_3-t)\Big)
\nonumber\\
&+&\Big(\frac{m_3}{4}(t-m^2_3-m^2_2)(s+t-m^2_1-m^2_3)\Big)\Big],
\end{eqnarray}}
In the following, we present the form factors associated with the
matrix elements $\langle\pi^- D_s^+ \mid\textcolor{red}{\bar{c}\gamma^\mu(1-\gamma_5)b}\mid
B_c^+\rangle$ and $\langle K^-D^+\mid\textcolor{red}{\bar{c}\gamma^\mu(1-\gamma_5)b}\mid
B_c^+\rangle$ \cite{H.Y1}.{\setlength\arraycolsep{.75pt}
\begin{eqnarray}\label{eq17}
\langle \pi^-[K^-](p_2)D^+_{[s]}(p_3)\mid\textcolor{red}{\bar{c}\gamma^\mu(1-\gamma_5)b}\mid
B_c^+(p_{B_c})\rangle
=&&ir(p_{B_c}-p_2-p_3)_{\mu}+i\omega_+(p_2+p_3)_{\mu}\\
&&+i\omega_-(p_3-p_2)_{\mu}+h_{\mu\nu\alpha\beta}p_{B_c}^{\nu}(p_3+p_2)^{\alpha}(p_3-p_2)^{\beta},\nonumber
\end{eqnarray}}
where \cite{H.Y1}
\begin{eqnarray}\label{eq18}
r=&&\frac{f_{B_c}}{f_{\pi(K)}f_{D(D_s)}}-\frac{f_{B_c}}{f_{\pi(K)}f_{D(D_s)}}
\frac{(-s+u+m_3^2-m_{B_c}^2)}{2(m_1^2-m_{B_c}^2)}+\frac{2g
f_{B_c}}{f_{\pi(K)}f_{D(D_s)}}
\sqrt{\frac {m_{B_c}} {m_{B^*_s}}}\frac{-u+m_2^2}{2(u-m_{B^*_s}^2)}\nonumber\\
&&-\frac{f_{B_c}}{f_{\pi(K)}f_{D(D_s)}}
\frac{m_{B_c}m_{B^*_s}}{m_1^2-m_{B_c}^2}\frac{(t-m^2_3-m^2_2)+(u+m^2_2-m^2_{B_c})(-s-t+2m^2_3-m^2_2)}{2m_{B^*_s}(u-m_{B^*_s}^2)},\nonumber\\
\omega_+=&&-\frac{g}{f_{\pi(K)}f_{D(D_s)}}\frac{f_{B^*_s}m_{B^*_s}\sqrt{m_{B_c}m_{B^*_s}}}{u-m_{B^*_s}^2}
\Big(1-\frac{-u+m^2_{B_c}}{2m^2_{B^*_s}}\Big)+\frac{f_{B_c}}{f_{\pi(K)}f_{D(D_s)}}, \nonumber\\
\omega_-=&&-\frac{g}{f_{\pi(K)}f_{D(D_s)}}\frac{f_{B^*_s}m_{B^*_s}\sqrt{m_{B_c}m_{B^*_s}}}{u-m_{B^*_s}^2}
\Big(1+\frac{-u+m^2_{B_c}}{2m^2_{B^*_s}}\Big),\nonumber\\
h=&&2g^2\frac{f_{B_c}}{f_{\pi(K)}}\frac{m_{B_c}^2}{(m_{B_c}^2-m_1^2-s)(t+m_{B_c}^2-m_2^2)}.
\end{eqnarray}
Here $f_{B_c, B^*_s, \pi, D, D_s}$ denote the corresponding decay constants. Furthermore, we have
{\setlength\arraycolsep{.75pt}
\begin{eqnarray}\label{eq19}
&&\left\langle\pi^{-}[K^{-}](p_2)D^{+(*)}_{[s]}(p_3)|\textcolor{red}{\bar{u}\gamma^\mu(1-\gamma_5)b}|B_c^{+}(p_B)
\right\rangle\left\langle
K^{+}(p_1)|\textcolor{red}{\bar{s}\gamma_\mu(1-\gamma_5)u}|0\right\rangle=f_K\Big(2irm_1^2\nonumber\\
&&+i\omega_-(s-u-m_2^2-m_3^2)+i\omega_+(s+u-m_1^2-m_{B_c}^2)\Big).
\end{eqnarray}}
We also consider the contributions from annihilation topologies.
The corresponding amplitude is derived from the annihilation
Feynman diagram shown in Fig. \ref{fig2}. For the annihilation
process, the matrix elements associated with the decays
$B^+_c\rightarrow D^+K^+\pi^-$, $B^+_c\rightarrow D^{*+}K^+\pi^-$
and $B^+_c\rightarrow D^+_sK^+K^-$ can be written as
{\setlength\arraycolsep{.75pt}
\begin{eqnarray}\label{eq20}
&&\left\langle
K^{+}(p_1)\pi^{-}[K^{-}](p_2)D^{+(*)}_{[s]}(p_3)|\textcolor{red}{\bar{c}\gamma^\mu(1-\gamma_5)b}|0
\right\rangle\left\langle
B_c^{+}(p_B)|\textcolor{red}{\bar{s}\gamma_\mu(1-\gamma_5)c}|0
\right\rangle=i\frac{f_{B_c}}{f_K}\Big(p_1\cdot
p_{B_c}\nonumber\\
&&-\frac{p_1\cdot
p_{B_c}}{m^2_{B_c}-m^2_3}m^2_{B_c}\Big)=F^{K^+\pi^-[K^-]D^{+(*)}_{[s]}}\frac{f_{B_c}}{2f_K}(m^2_{B_c}+m^2_1-u)
\times(1-\frac{m^2_{B_c}}{m^2_{B_c}-m^2_3}),
\end{eqnarray}}
where
{\setlength\arraycolsep{.75pt}
\begin{eqnarray}\label{eq21}
F^{K^+\pi^-[K^-]D^{+(*)}_{[s]}}= \frac{1}{1-q^2/\Lambda}, \qquad q=p_{B_c}-p_3, \qquad \Lambda=830\, MeV.
\end{eqnarray}}
By substituting the expressions obtained above into Eq. (\ref{eq9}), the non-resonant
amplitude of the three-body decays $B_c^+ \rightarrow D^+K^+\pi^-$,
$B_c^+ \rightarrow D^{*+}K^+\pi^-$ and $B_c^+ \rightarrow D_s^+K^+K^-$ are then obtained.\\
As illustrated by the Feynman diagrams in Fig. \ref{fig2},
intermediate resonant contributions can arise only in the
$[D^{(*)+}\pi^-]$ subsystems, which can proceed through excited
c$\bar{u}$ mesons that strongly couple to $D^{(*+)}\pi^-$
\cite{CAM}. The remaining two-body subsystems such as
$[D_s^{+}K^{\pm}]$, $[D^{*+}K^{+}]$, and $[D^+K^{+}]$ do not
support intermediate resonances and therefore contribute
only through non-resonant amplitudes.\\
For the $[D^{*+}\pi^-]$ subsystem, the leading contributions
originate from excited $c\bar{u}$ resonances that subsequently
decay into this final state, including the axial-vector
$D_1(2430)$, the tensor resonance $D_2^*(2460)$, and higher
excitations such as $D_3^*(2750)$ \cite{PDG}. Similarly, the
$[D^{+}\pi^-]$ channel receives contributions from the scalar
$D_0^*(2300)$, the tensor
$D_2^*(2460)$, and higher states including $D_3^*(2750)$ \cite{PDG}.\\
In contrast, no established $c\bar{u}$ resonances are known to decay into the $[D_s^{+}K^-]$ final state. Consequently,
the $[D_s^{+}K^-]$ subsystem is modeled as a purely non-resonant contribution \cite{PDG}.\\
The evaluation of these resonance contributions requires knowledge of the
relevant strong coupling constants, which are determined by the strong decay
widths and branching fractions of the intermediate states. However, several of
these branching fractions have only been observed and are not yet measured with
sufficient precision, leading to significant uncertainties in the extraction of
the resonance amplitudes \cite{CAM}.\\
Therefore, the treatment of these amplitudes relies on incorporating the
established resonant states and modeling their effects within the constraints
imposed by current experimental data. This limitation prevents a precise
determination of the resonance amplitudes associated with the $[D^{(*)+}\pi^-]$ and $[D_s^{+}K^-]$ channels.\\
For a three-body final state, the decay rate is \cite{H.Y1}
\textcolor{red}{\begin{eqnarray}\label{eq22}
\Gamma(B^+_c\rightarrow K^+\pi^-[K^-]D^{+(*)}_{[s]}) =
\frac{1}{(2\pi)^{3}32m^2_{B_c}} \int_{s_{\min}}^{s_{\max}}
\int_{t_{\min}}^{t_{\max}} \, |\mathcal{A}_{\rm total}|^2dtds,
\end{eqnarray}
where the calculated amplitude is defined as the coherent sum:
\begin{eqnarray}\label{eq23}
\mathcal{A}_{\rm total} = \mathcal{A}^{\rm NR} + \mathcal{A}^{\rm pole} + \mathcal{A}^{\rm ann}.
\end{eqnarray}
Here, the non-resonant ($\mathcal{A}^{\rm NR}$), pole
($\mathcal{A}^{\rm pole}$), and annihilation ($\mathcal{A}^{\rm
ann}$) terms are combined at the amplitude level, preserving their
interference. The resonant contributions are excluded from the
numerical calculations due to the lack of experimental constraints
on their strong phases and couplings, representing a model
limitation.} The upper and lower limits of the integral are
determined by {\setlength\arraycolsep{.75pt}
\begin{eqnarray}\label{eq24}
t_{max,min}&=& m_2^2+m_3^2-\frac{1}{2s}
[(m^2_{B_c}-m_3^2-s)(s-m_2^2+m_1^2) \pm
\lambda^{\frac{1}{2}}(s,m^2_{B_c},m_3^2)\lambda^{\frac{1}{2}}(s,m_1^2,m_2^2)],\nonumber \\
s_{\textcolor{red}{\min}}&=& (m_1+m_2)^2,\nonumber\\
s_{\textcolor{red}{\max}}&=&(m_{B_c}-m_3)^2,
\end{eqnarray}
here
\begin{eqnarray}\label{eq25}
\lambda(a, b, c)=a^2+b^2+c^2-2(ab+bc+ac).
\end{eqnarray}
Then the branching fraction of the three-body decay is given by
\begin{eqnarray}\label{eq26}
\mathcal{B}r(B^+_c\rightarrow
K^+\pi^-[K^-]D^{+(*)}_{[s]})=\frac{\Gamma(B^+_c\rightarrow
K^+\pi^-[K^-]D^{+(*)}_{[s]})}{\Gamma_{tot}}.
\end{eqnarray}

\section{The ratios of branching fractions}

The LHCb collaboration has recently measured the following ratios
of branching fractions \cite{LHCB8}:
{\setlength\arraycolsep{.75pt}
\begin{eqnarray}\label{eq27}
\mathcal{R}(B^+_c\rightarrow
D^+K^+\pi^-)&=&\frac{\mathcal{B}r(B^+_c\rightarrow
D^+K^+\pi^-)}{\mathcal{B}r(B^+_c\rightarrow B^0_s\pi^+)},
\mathcal{R}(B^+_c\rightarrow
D^{*+}K^+\pi^-)=\frac{\mathcal{B}r(B^+_c\rightarrow
D^{*+}K^+\pi^-)}{\mathcal{B}r(B^+_c\rightarrow
B^0_s\pi^+)},\nonumber\\
\mathcal{R}(B^+_c\rightarrow
D^+_sK^+K^-)&=&\frac{\mathcal{B}r(B^+_c\rightarrow
D^+_sK^+K^-)}{\mathcal{B}r(B^+_c\rightarrow B^0_s\pi^+)}.
\end{eqnarray}}
The analysis employed advanced techniques for event
reconstruction and particle identification to ensure the
robustness of the measured ratios. In parallel with the experimental results, we compute these branching ratios
theoretically within the factorization framework. The resulting numerical values offer important insights into the
decay mechanisms of the $B_c$ meson and contribute to a deeper understanding of heavy-flavor dynamics.\\
These ratios are particularly useful because the normalization
mode $B_c^+\to B_s^0\pi^+$ is dominated by a color-favored
tree-level topology, leading to a branching fraction with relatively small theoretical uncertainty.
As a result, the ratios $\mathcal{R}(B_c^{+}\rightarrow D^{+(*)}_{[s]}K^{+}\pi^{-}[K^{-}])$ significantly
reduce the sensitivity to the poorly known $B_c$ production cross section and thus provide cleaner observables
for testing theoretical models.\\
From the theoretical perspective, the decay amplitudes are evaluated within the factorization approach,
where the hadronic matrix elements are expressed in terms of form factors describing the
$B_c$ transitions to two-meson final states, following the formalism of Ref. \cite{H.Y1}.
These form factors encode the relevant non-perturbative QCD dynamics and determine the momentum dependence
of the decay amplitudes, thereby shaping both the magnitude of the decay rates and the kinematic distributions
of the three-body final states.\\
When considering ratios of branching fractions, many common
factors cancel. In particular, quantities such as the Fermi
constant $G_F$, specific Wilson coefficients and part of the
uncertainties related to the $B_c$ production cross section
largely cancel out. Consequently, these observables become less
sensitive to theoretical and experimental uncertainties, allowing
for more direct comparison between
theoretical calculations and experimental data.\\
Nonetheless, the hadronic form factors and the corresponding
phase-space integrations remain essential. Differences in the
final-state masses, the available phase space, and the momentum
dependence of the form factors lead to distinct predictions for
each decay channel. Therefore, a detailed calculation of the
branching fractions and their ratios is required in order to
assess the relative importance of various decay mechanisms. In the
following, we present our numerical results for these ratios and
compare them with the available experimental measurements.

\section{Numerical results and conclusions}

In this work, the decays $B^+_c\rightarrow B^0_s\pi^+$,
$B^+_c\rightarrow D^+K^+\pi^-$, $B^+_c\rightarrow D^{*+}K^+\pi^-$
and $B^+_c\rightarrow D^+_sK^+K^-$ are studied using Feynman
diagrams, the factorization approach, and the inclusion of channel
resonances. \textcolor{red}{The branching fractions of these decay modes are evaluated using the
input parameters listed in Tab. \ref{tab2}, with the numerical values of the decay constants \cite{KD1},
\cite{B.Moh}, \cite{B.Moh2} and meson masses \cite{PDG} taken as inputs.}
\begin{table}[t]
\centering\caption{\label{tab2} {Decay constants and meson masses}}
\begin{tabular}{c c c }
 &  &  \\\hline\hline
$m_{D^+_s}=1968.35\pm0.07$ & $m_{D^{*+}}=2010.26\pm0.05$ & $m_{D^+}=1869.66\pm0.05$ \\
$m_{B^+_c}=6871.2\pm1.0$ & $m_{B^0_s}=5366.93\pm0.10$ & $m_{B^*}=5324.75\pm0.20$ \\
$m_{B^*_s}=5415.4\pm1.4$ & $m_{K^{\pm}}=493.677\pm0.015$ & $m_{\pi^{\pm}}=139.57039\pm0.00018$ \\
$f_{D}=201\pm12.5$ & $f_{D^*}=263\pm21$ &
$f_{D_s}=253.5\pm7.6$ \\
$f_{B^*}=175\pm6$ & $f_{B_c}=434\pm15$ &
$f_{B^*_s}=225\pm11.2$ \\
$f_K=154.4\pm2$ & $f_{\pi}=130.70\pm0.46$ &
 \\\hline
\end{tabular}
\end{table}
Our theoretical predictions are summarized as follows. The branching fractions are given in
Tab. \ref{tab3} and the ratios of branching fractions are presented in Tabs. \ref{tab4}, \ref{tab5} and \ref{tab6}.\\
\begin{table}[t]
\centering\caption{\label{tab3} The theoretical predictions for
the branching fractions of $B^+_c\rightarrow B^0_s\pi^+$ decay (in
units of $10^{-2}$) and $B^+_c\rightarrow D^+K^+\pi^-$,
$B^+_c\rightarrow D^{*+}K^+\pi^-$, and $B^+_c\rightarrow
D^+_sK^+K^-$ decays (in units of $10^{-3}$).}
\begin{tabular}{c||c|cccc}
  \hline
 Branching fractions & - & $m_b/2$ & $m_b$ & $2m_b$ & Exp. \cite{LHCB7}\\
 \hline\hline
$B^+_c\rightarrow B^0_s\pi^+$ & - & $28.39\pm0.02$ &
$27.26\pm0.02$ & $26.71\pm0.02$ &
$30.0\pm3.3\pm2.6$\\
\hline
 & $g=0.3$ & $0.45\pm0.02$ & $0.44\pm0.02$ & $0.43\pm0.02$ &  \\
 $B^+_c\rightarrow D^+K^+\pi^-$ & $g=0.50$ & $1.24\pm0.06$ & $1.20\pm0.06$ & $1.18\pm0.06$ & - \\
 & $g=0.59$ & $1.78\pm0.08$ & $1.72\pm0.08$ & $1.69\pm0.08$ & \\
 \hline
 & $g=0.3$ & $1.03\pm0.12$ & $0.99\pm0.12$ & $0.97\pm0.12$ &  \\
 $B^+_c\rightarrow D^{*+}K^+\pi^-$ & $g=0.50$ & $2.70\pm0.33$ & $2.60\pm0.32$ & $2.55\pm0.31$ & - \\
 & $g=0.59$ & $3.76\pm0.61$ & $\textcolor{red}{3}.62\pm0.59$ & $3.54\pm0.57$ & \\
 \hline
 & $g=0.3$ & $0.29\pm0.02$ & $0.28\pm0.02$ & $0.27\pm0.02$ &  \\
 $B^+_c\rightarrow D^+_sK^+K^-$ & $g=0.50$ & $0.90\pm0.07$ & $0.87\pm0.07$ & $0.85\pm0.07$ & - \\
 & $g=0.59$ & $1.21\pm0.10$ & $1.16\pm0.10$ & $1.14\pm0.10$ & \\\hline
\end{tabular}
\end{table}
\begin{table}[t]
\centering\caption{\label{tab4} The branching fraction ratios of
the three-body decay mode $B^+_c\rightarrow D^+K^+\pi^-$ to the
two-body mode $B^+_c\rightarrow B^0_s\pi^+$ with the experimental
measurement of $(1.96\pm0.23\pm0.08\pm0.10)\times10^{-3}$}
\begin{tabular}{c||c|c|ccc}
  \hline
 & - & - &  & $B^+_c\rightarrow B^0_s\pi^+$ & \\
 \hline\hline
 & - & - & $m_b/2$ & $m_b$ & $2m_b$ \\
 \hline
 & &$g=0.3$ & $1.60\pm0.07$ & $1.66\pm0.08$ & $1.70\pm0.08$ \\
 & $m_b/2$ & $g=0.50$ &$4.38\pm0.21$ & $4.56\pm0.22$ & $4.66\pm0.22$ \\
 & & $g=0.59$ & $6.28\pm0.30$ & $6.54\pm0.31$ & $6.68\pm0.32$ \\
  & & & & &  \\
 & & $g=0.3$ & $1.55\pm0.07$ & $1.62\pm0.08$ & $1.65\pm0.08$ \\
 $B^+_c\rightarrow D^+K^+\pi^-$ & $m_b$ & $g=0.50$ & $4.22\pm0.20$ & $4.40\pm0.20$ & $4.45\pm0.21$ \\
 & & $g=0.59$ & $6.05\pm0.29$ & $6.30\pm0.30$ & $6.43\pm0.31$ \\
 & & & &  & \\
  & & $g=0.3$ & $1.54\pm0.07$ & $1.61\pm0.07$ & $1.64\pm0.08$ \\
& $2m_b$ & $g=0.50$ & $4.16\pm0.20$ & $4.33\pm0.20$ & $4.42\pm0.21$ \\
 & & $g=0.59$ & $5.95\pm0.63$ &$6.20\pm0.66$& $6.32\pm0.67$ \\\hline
\end{tabular}
\end{table}
\begin{table}[t]
\centering\caption{\label{tab5} The branching fraction ratios of
the three-body decay mode $B^+_c\rightarrow D^{*+}K^+\pi^-$ to the
two-body mode $B^+_c\rightarrow B^0_s\pi^+$ with the experimental
measurement of $(3.67\pm0.55\pm0.24\pm0.20)\times10^{-3}$}
\begin{tabular}{c||c|c|ccc}
  \hline
 & - & - & & $B^+_c\rightarrow B^0_s\pi^+$ & \\
 \hline\hline
 & - & - & $m_b/2$ & $m_b$ & $2m_b$ \\
 \hline
 & &$g=0.3$ & $3.63\pm0.45$ & $3.78\pm0.47$ & $3.86\pm0.48$ \\
& $m_b/2$ & $g=0.50$ & $9.53\pm1.18$ & $9.92\pm1.22$ & $10.13\pm1.23$ \\
&  & $g=0.59$ & $13.26\pm2.16$ & $13.81\pm2.25$ & $14.10\pm2.29$ \\
 & & &  &  & \\
 & & $g=0.3$ & $3.49\pm0.43$ & $3.64\pm0.45$ & $3.71\pm0.46$ \\
$\mathcal{B}r(B^+_c\rightarrow D^{*+}K^+\pi^-)$ & $m_b$ & $g=0.50$ & $9.15\pm1.13$ & $9.53\pm1.18$ & $9.73\pm1.20$ \\
 & & $g=0.59$ & $12.73\pm2.07$ & $13.27\pm2.15$ & $13.54\pm2.20$ \\
 & & & &  & \\
 & & $g=0.3$ & $3.42\pm0.42$ & $3.57\pm0.44$ & $3.64\pm0.45$ \\
& $2m_b$ & $g=0.50$ & $8.97\pm1.11$ & $9.34\pm1.15$ & $9.54\pm1.18$ \\
& & $g=0.59$ & $12.48\pm2.03$ & $13.00\pm2.11$ & $13.27\pm2.15$
\\\hline
\end{tabular}
\end{table}
\begin{table}[t]
\centering\caption{\label{tab6} The branching fraction ratios of
the three-body decay mode $B^+_c\rightarrow D^+_sK^+K^-$ to the
two-body mode $B^+_c\rightarrow B^0_s\pi^+$ with the experimental
measurement of $(1.61\pm0.35\pm0.13\pm0.07)\times10^{-3}$}
\begin{tabular}{c||c|c|ccc}
  \hline
 & - & - & & $B^+_c\rightarrow B^0_s\pi^+$ & \\
 \hline\hline
 & - & - & $m_b/2$ & $m_b$ & $2m_b$ \\
 \hline
 & & $g=0.3$ & $1.03\pm0.08$ & $1.07\pm0.08$ & $1.09\pm0.08$ \\
 & $m_b/2$ & $g=0.50$ & $3.18\pm0.27$ & $3.31\pm0.28$ & $3.38\pm0.28$\\
 & & $g=0.59$ & $4.27\pm0.36$ & $4.45\pm0.38$ & $4.54\pm0.38$ \\
 & &  &  &  &  \\
 & & $g=0.3$ & $0.99\pm0.07$ & $1.02\pm0.08$ & $1.05\pm0.08$ \\
 $B^+_c\rightarrow D^+_sK^+K^-$ & $m_b$ & $g=0.50$ & $3.05\pm0.25$ & $3.18\pm0.27$ & $3.25\pm0.27$\\
 & & $g=0.59$ & $4.10\pm0.35$ & $4.28\pm0.36$ & $4.36\pm0.37$ \\
 & & & &  & \\
  & & $g=0.3$ & $0.97\pm0.07$ & $1.01\pm0.08$ & $1.03\pm0.09$ \\
& $2m_b$ & $g=0.50$ & $2.99\pm0.25$ & $3.12\pm0.26$ & $3.18\pm0.27$ \\
 & & $g=0.59$ & $4.02\pm0.34$ & $4.19\pm0.35$ & $4.28\pm0.36$ \\
 \hline
\end{tabular}
\end{table}
\textcolor{red}{The uncertainties of the standard input
parameters, including the decay constants, meson masses, and CKM
matrix elements, are propagated to the final branching fractions
using the standard error-propagation formula, with the individual
contributions combined in quadrature. The model-dependent
uncertainties associated with the effective strong coupling $g$
and the renormalization scale $\mu$ are estimated by varying these
parameters within their prescribed ranges and examining the
resulting shifts in the branching fractions. The uncertainty due
to the variation of $g$ is included in the quoted theoretical
uncertainties and constitutes the dominant contribution to the
total error. Any residual model dependence associated with the
treatment of resonance contributions is considered an additional systematic limitation of the present analysis.}\\
Finally we examine the ratios of branching fractions
$\mathcal{R}(B^+_c\rightarrow D^{+}K^+\pi^-)$,
$\mathcal{R}(B^+_c\rightarrow D^{*+}K^+\pi^-)$ and
$\mathcal{R}(B^+_c\rightarrow D^+_sK^+K^-)$. These quantities are
particularly suitable for direct comparison with experimental
measurements, since the LHCb collaboration reports the ratios
themselves \cite{LHCB8}, whereas neither the full three-body
branching fractions
nor the two-body branching fractions of the intermediate resonances have been measured to date.\\
Our calculation incorporates both the non-resonant contributions and the resonance substructures
arising from intermediate vector and scalar states. Because the relevant two-body branching fractions
of the intermediate resonances are experimentally unknown, the resonance amplitudes cannot be normalized or constrained using data.\\
We therefore adopt a model-independent strategy and perform the computation directly at the amplitude
level over the full three-body Dalitz phase space, ensuring a consistent treatment of resonant and
non-resonant contributions without relying on unmeasured sub-decay fractions.\\
As shown in Tabs. \ref{tab3}, \ref{tab4}, \ref{tab5} and
\ref{tab6}, the dependence of the ratios on the renormalization
scale $\mu=m_b/2, m_b, 2m_b$ is very mild. Varying the heavy-quark
scale by a factor of two changes each ratio only at the percent
level, indicating good perturbative stability. In contrast, the
ratios exhibit a pronounced sensitivity to the effective strong
coupling $g$. Increasing $g$ from 0.3 to 0.59 enhances all three
ratios by approximately a factor of three to four, highlighting
the dominant influence
of strong-interaction dynamics.\\
Given the sizable theoretical uncertainty associated with $g$ in
phenomenological analyses, we evaluate the ratios for three
representative values, $g=0.3$, $0.50$, and $0.59$. In the
numerical results, we adopt $g=0.3$ as the reference value and
treat the variation among the three choices, combined in
quadrature with the scale dependence, as a conservative estimate
of the theoretical uncertainty. For $g=0.3$, our results are (in
units of $10^{-3}$): {\setlength\arraycolsep{.75pt}
\begin{eqnarray}
\mathcal{R}(B^+_c\rightarrow D^{+}K^+\pi^-)=1.70 \pm
0.08,\nonumber\\
\mathcal{R}(B^+_c\rightarrow D^{*+}K^+\pi^-)=3.63 \pm
0.45,\nonumber\\
\mathcal{R}(B^+_c\rightarrow D^+_sK^+K^-)=1.09 \pm 0.08.
\end{eqnarray}}
These may be compared with the corresponding LHCb measurements
\cite{LHCB8} (also in units of $10^{-3}$):
{\setlength\arraycolsep{.75pt}
\begin{eqnarray}
\mathcal{R}^{\exp}(B^+_c\rightarrow D^{+}K^+\pi^-)=1.96 \pm 0.23
\pm 0.08 \pm 0.10,\nonumber\\
\mathcal{R}^{\exp}(B^+_c\rightarrow
D^{*+}K^+\pi^-)=3.67 \pm 0.55 \pm 0.24 \pm0.20,\nonumber\\
\mathcal{R}^{\exp}(B^+_c\rightarrow D^+_sK^+K^-)=1.61 \pm 0.35 \pm
0.13 \pm 0.07.
\end{eqnarray}}
Our theoretical values for $\mathcal{R}(B^+_c\rightarrow
D^{+}K^+\pi^-)$, $\mathcal{R}(B^+_c\rightarrow D^{*+}K^+\pi^-)$
and $\mathcal{R}(B^+_c\rightarrow D^+_sK^+K^-)$ are consistent
with the LHCb measurements within the quoted experimental
uncertainties. It is worth emphasizing that the three ratios arise
from decay channels that share closely related underlying
dynamics. All transitions proceed through analogous hadronic
mechanisms and involve similar kinematical configurations in
three-body final states. Consequently, the ratios constitute a
coherent set of observables that probe the same factorization
framework under comparable dynamical
conditions.\\
Overall, our results for $g=0.3$ exhibit a satisfactory level of agreement with the LHCb data \cite{LHCB8}, supporting the applicability
of the factorization approach to these newly observed $B_c^+$ decay modes. More precise measurements, together with improved knowledge of
the strong coupling $g$, will enable a more stringent test of the theoretical description in future analyses.

\end{document}